\documentclass[pdflatex, sn-mathphys]{sn_arXiv}

\jyear{2021}%

\theoremstyle{thmstyleone}%
\theoremstyle{thmstyletwo}%

\theoremstyle{thmstylethree}%

\begin{document}

\title[Spatial Structure of City Population Growth]{Spatial Structure of City Population Growth}



\author[1]{\fnm{Sandro M.} \sur{Reia}}

\author[1,2]{\fnm{P. Suresh C.} \sur{Rao}}

\author[3,4]{\fnm{Marc} \sur{Barthelemy}}

\author*[1]{\fnm{Satish V.} \sur{Ukkusuri}}\email{sukkusur@purdue.edu}

\affil*[1]{\orgdiv{Lyles School of Civil Engineering}, \orgname{Purdue University}, \orgaddress{\street{550 Stadium Mall Avenue}, \city{West Lafayette}, \postcode{47907}, \state{Indiana}, \country{USA}}}

\affil[2]{\orgdiv{Agronomy Department}, \orgname{Purdue University}, \orgaddress{\street{915 W State Street}, \city{West Lafayette}, \postcode{47907}, \state{Indiana}, \country{USA}}}

\affil[3]{\orgdiv{Institut de Physique Théorique}, \orgname{CEA}, \orgaddress{\street{CNRS-URA 2306}, \city{Gif-surYvette}, \postcode{91191}, \country{France}}}

\affil[4]{\orgdiv{Centre d’Analyse et de Mathématique Sociales}, \orgname{CNRS/EHESS}, \orgaddress{\city{Paris}, \postcode{75006}, \country{France}}}


\abstract{
We show here that \textcolor{black}{population growth, resolved at the county level, is spatially heterogeneous both among and within the U.S. metropolitan statistical areas}. 
Our analysis of data for over 3,100 U.S. counties reveals that annual population flows, resulting from \textcolor{black}{domestic migration} during the 2015 - 2019 period, are much larger than natural demographic growth, and are primarily responsible for this heterogeneous growth. More precisely, we show that intra-city \textcolor{black}{flows} are generally along a negative population density gradient, while inter-city \textcolor{black}{flows} are concentrated in high-density core areas. 
Intra-city \textcolor{black}{flows} are anisotropic and generally directed towards external counties of cities, driving asymmetrical urban sprawl. 
Such \textcolor{black}{domestic migration} dynamics are also responsible for tempering local population shocks by redistributing inflows within a given city. This “spill-over” effect leads to a smoother population dynamics at the county level, in contrast to that observed at the city level. 
Understanding the spatial structure of \textcolor{black}{domestic migration} flows is a key ingredient for analyzing their drivers and consequences, thus representing a crucial knowledge for urban policy makers and planners. 
}

\keywords{Metropolitan Statistical Areas, Internal Migration, Urbanization, Relocation Flows, City Growth}



\maketitle



\section{Introduction}\label{sec1}

Research on city population growth has a long history with statistical regularities among the cities, as identified in early seminal works by Auerbach \cite{auerbach1913gesetz} and later by Zipf \cite{george1949zipf}. Random demographic growth  was generally considered \cite{gabaix1999zipf, gibrat1931inegalites} as the main source of the population growth dynamics of cities. However, recently \cite{verbavatz2020growth, bettencourt2020demography} city population growth was shown to result from a combination of random demographic growth and, more importantly, inter-city \textcolor{black}{flows from domestic migration} that are broadly distributed according to a power law \cite{verbavatz2020growth}. These flows are triggered by socioeconomic changes and can dramatically alter the trajectory of the population growth of a city  \cite{monras2018economic, vsveda2016behind}. 

\textcolor{black}{Inter-city flows from domestic migration} play a crucial role in the evolution of the system of cities at a country scale and its analysis is fundamental to understand the \textcolor{black}{temporal and spatial evolution of cities}.
More specifically, the structure of household \textcolor{black}{domestic migration} provides insights on \textcolor{black}{regions} that are more likely to grow, which is usually accompanied by various externalities, such as traffic congestion \cite{downs2005still, hymel2009does, sweet2011does}, air pollution \cite{darccin2014association, portney2013taking} and socioeconomic inequality \cite{musterd2017socioeconomic}. 
Extreme flows lead to unprecedented population growth that is usually more expansive than compact \cite{seto2011meta, wu2012land, schneider2014expansion}, \textcolor{black}{thus} understanding the structure of \textcolor{black}{domestic migration} flows at intra- and inter-city scales help to plan for various unforeseen problems and to devise mitigation strategies. This is particularly important and well-known for the suburban and fringe area urbanization  \cite{lichter2021rural}, which have their own planning challenges and peculiarities \cite{phelps2021city}.

The dynamics of city population growth are usually studied at the city-level, neglecting the intra-city spatial structure of \textcolor{black}{migratory} flows. 
Spatial heterogeneity of cities is well known, evident in consistent patterns, among others, nonlinear decrease in population density with increasing distance from the dense urban core \cite{newling1969spatial}, fractal urban morphology \cite{batty1991cities}, spatial structure of urban heat islets \cite{shreevastava2019}. 
Other urban studies focused on emergence of inequalities among neighborhoods and infrastructure development \cite{vsveda2016behind, hao2020effect}. In line with these studies, we consider here the spatial heterogeneity in \textcolor{black}{city} population growth by conducting an  analysis of  data for \textcolor{black}{domestic migration} flows in the U.S. \textcolor{black}{using the most recent American Community Survey (ACS) 5-year county-to-county migration flow files, which are available for the time period $2015$ - $2019$ \cite{ACS}}. In particular, we focus on the origin and destination counties \textcolor{black}{of the domestic migration flows}, revealing spatial variations \textcolor{black}{of components of} city population growth at the county level and heterogeneous growth within metropolitan statistical areas.
\textcolor{black}{For this reason, in the following discussion we interchangeably use the terms city and metropolitan statisical area (MSA).
Also, population flows between U.S cities (inter-city) and within U.S. cities (intra-city), among counties within the same city, are examined.
}

\textcolor{black}{There are over $380$ metropolitan statisical areas in the U.S.} \cite{census}, each composed of one or more counties accounting for about $86\%$ of the U.S. population.
We analyze in detail the importance of both inter- and intra-city flows in the \textcolor{black}{city population} growth dynamics. Given that the U.S. is highly urbanized, two other components play a much smaller role in the population growth dynamics of MSAs: \textcolor{black}{flows between metro and micropolitan statistical areas, and those between metro and non-statistical areas.}
Therefore, our goal here is to examine generalized patterns across cities, despite their specific differences, rather than focusing on topological properties of flow networks \cite{slater2008hubs, slater1976hierarchical}.

\begin{figure}[t!]
\centering
\includegraphics[width=0.65\linewidth]{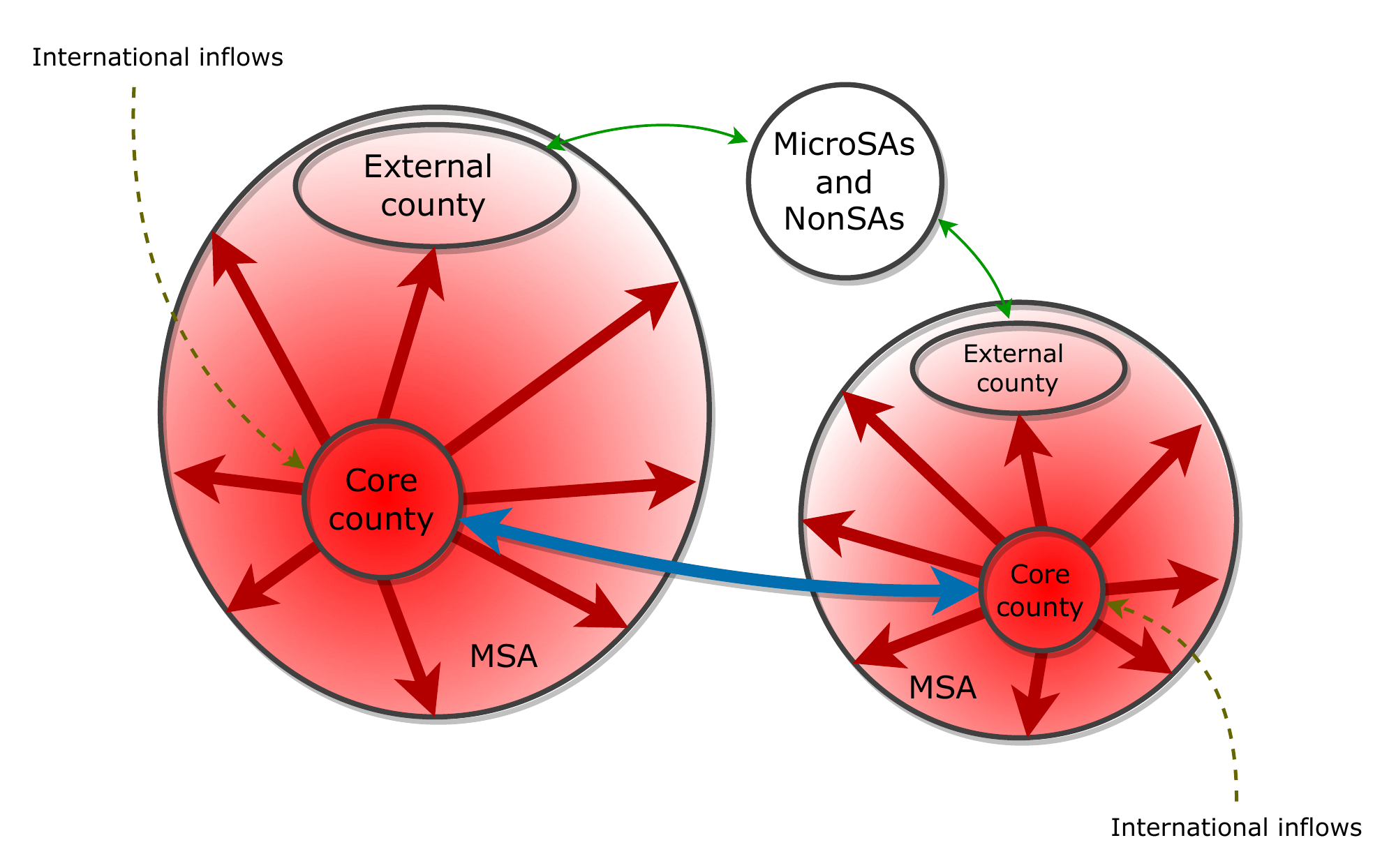}
\caption{\textcolor{black}{Schematic representation of the dominant migratory trends that contribute to the heterogeneous population growth of cities}. 
Core counties are more likely to receive inflows from core counties of other cities than from external counties (blue arrows).
Flows to and from \textcolor{black}{micro and non-statistical} areas are more likely to be found at the external counties of a city (green arrows).
Intra-city flows (red arrows) indicate vectors of redistribution of people within the city, and have an outwards radial direction: people move from central counties with larger population density to external counties with lower densities. 
\textcolor{black}{International inflows (yellow dashed arrows), which scale superlinearly with city population, are more likely to be directed to the core counties of large cities.}
The resulting spatial heterogeneity is depicted by the background color, in which the red intensity is proportional to the population density. 
The width of the arrows \textcolor{black}{are proportional to the intensity of the flows}.}
\label{diagram}
\end{figure}

Our analysis provides a spatial and statistical structure of population flows resulting from \textcolor{black}{domestic migration} between and within cities, and helps to understand the heterogeneous spatial \textcolor{black}{population} growth of cities. Here, we will refer to these \textcolor{black}{domestic migration flows} as inflows or outflows, and netflows (inflows-outflows). 
We show that \textcolor{black}{inter-city} flows are more likely to occur between \textcolor{black}{core counties (the core county has the highest population density in a city)}, and \textcolor{black}{intra-city} flows are more likely to follow an outward radial direction (i.e., there is a trend towards exterior and lower density counties). 
Moreover, flows to/from \textcolor{black}{micro areas} and rural counties are more likely to happen at the external \textcolor{black}{regions} of cities (see Fig. \ref{diagram}).

\section{\textcolor{black}{Overview of U.S. domestic migration flows}}\label{sec2}


\begin{figure*}[t]
\centering
\includegraphics[width = 1.0\textwidth]{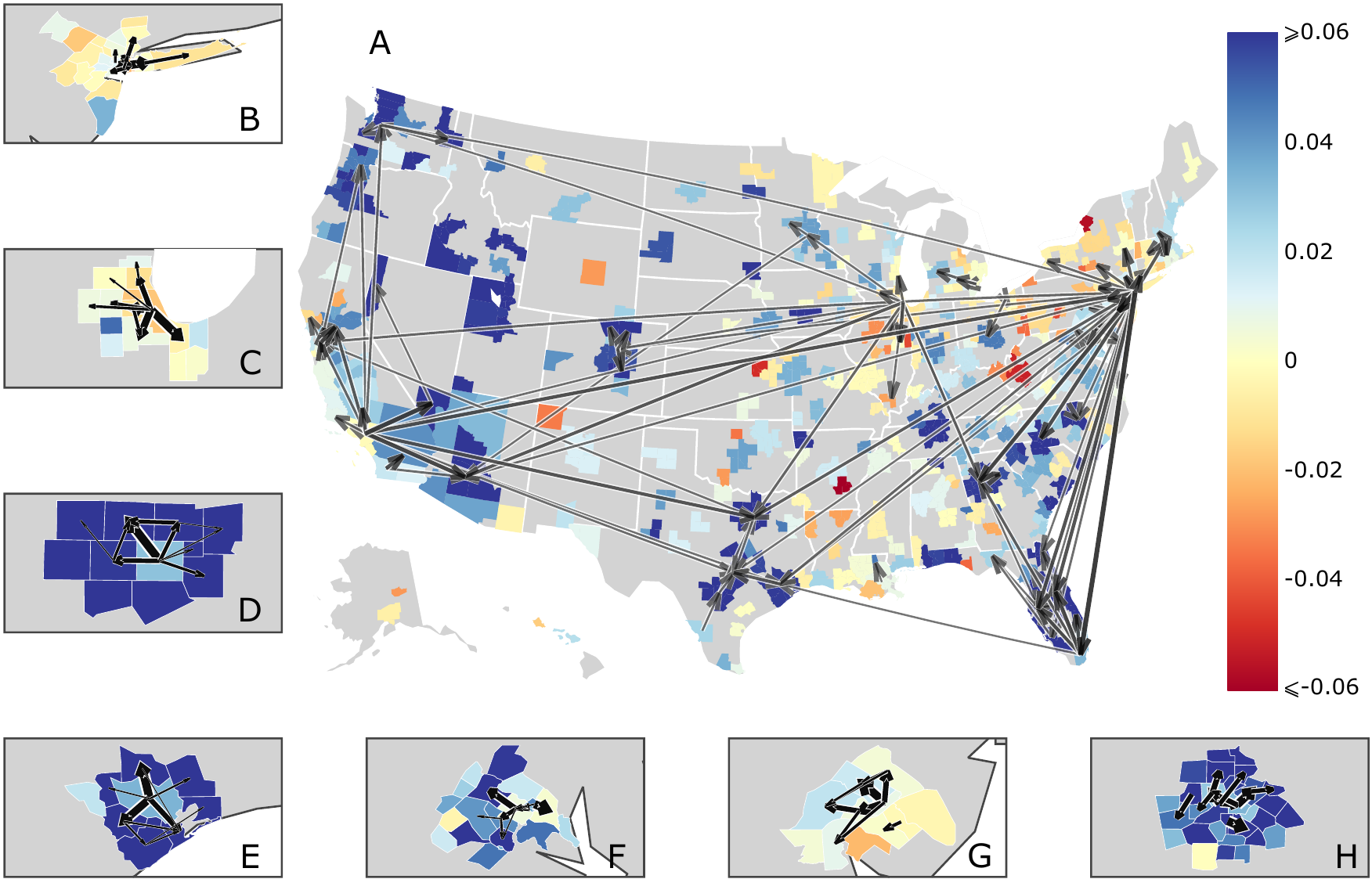}
\caption{Heterogeneity of inter- and intra-\textcolor{black}{city netflows}. The map (A) suggests that \textcolor{black}{the domestic redistribution of people} between different U.S. \textcolor{black}{metro areas} are non-uniform: the black arrows, indicating the direction of the most intense inter-\textcolor{black}{city} \textcolor{black}{netflows} (higher than $2,000$ people per year), reveal migration trends from northern and eastern cities to western and southern regions. 
Cities (composed of one or more counties) are colored according to the \textcolor{black}{relative growth (viz. population growth adjusted by population)} of the whole MSA during the 2015 - 2019 period, and the black intensity and the thickness of the arrows are proportional to the \textcolor{black}{netflows}.
Panels (B-H), which are close-up of New York (B), Chicago (C), Dallas (D), Houston (E), Washington D.C. (F), Philadelphia (G), Atlanta (H), suggest that the most intense intra-\textcolor{black}{city} \textcolor{black}{netflows} are oriented radially outwards: people are moving from core to external counties. 
Here, counties are colored according to their relative growth in the 2015 - 2019 period and the width of the arrows is proportional to the \textcolor{black}{netflows between origin and destination counties.}}
\label{map_flows}
\end{figure*}

\begin{figure}[h!] 
\centering
\includegraphics{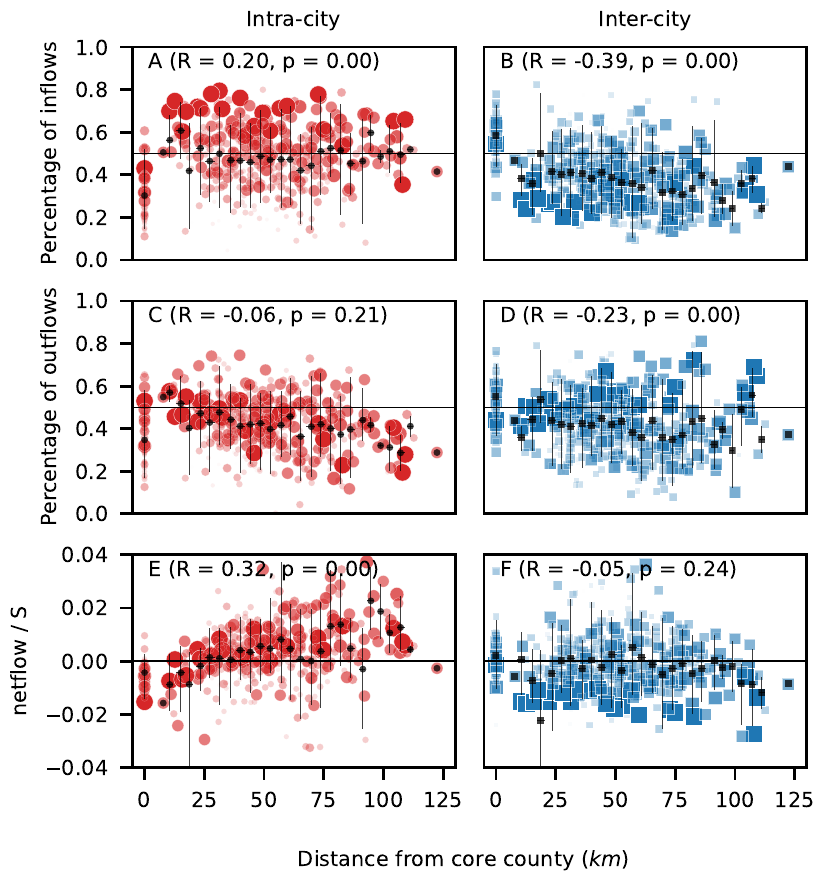}
\caption{Roles of intra- and inter-\textcolor{black}{city} flows in driving the heterogeneous \textcolor{black}{population growth} of cities. We define the core county as the one with the highest population density, and we plot the percentage of inflows due to intra- (A) and inter-\textcolor{black}{city} flows (B) of each county within a city as a function of its distance to the core county.
The percentage of outflows due to intra- and inter-\textcolor{black}{city} flows are shown in (C) and (D), respectively. 
The positive correlation of the relative growth with distance due to intra-\textcolor{black}{city} flows in (E), along with the lack of correlation due to inter-\textcolor{black}{city} flows in (F), indicates that intra-\textcolor{black}{city} flows are mainly responsible for increasing the population in the external \textcolor{black}{regions} of cities. 
The sizes of red circles and blue squares are proportional to the city population.
The black dots and the error bars indicate the mean and the $90\%$ interval, respectively, of the counties within the corresponding bin.
}
\label{flows_with_distance}
\end{figure}

\textcolor{black}{
The most recent ACS county-to-county flow dataset \cite{ACS} reports that about $45.6$ million people migrated in the U.S. per year during the period $2015$ - $2019$, which corresponds to $14.2\%$ of the U.S. population.
Approximately $43.5$ million of annual moves corresponded to domestic migration (moves within the U.S. \cite{koerber2007domestic}), while $2.1$ million accounted for inflows of individuals from other countries (viz. international immigration).
}

\textcolor{black}{With respect to domestic migration, $25.7$ million of people per year migrated within the same county, thus showing that the highest share of domestic flows ($59\%$) is intra-county.
Annually, about $10.4$ people moved between different counties within the same state, thus intra-state flows account for $24\%$ of the domestic migration
}(Supplementary Fig. 1), mainly driven by the search for more affordable housing, better jobs, and for family reasons such as change in marital status \cite{Frost2020}.
\textcolor{black}{
Long distance moves, captured by inter-state flows, represent the remaining $17\%$ of domestic flows, which comprises about $7.5$ million moves per year.
}


\textcolor{black}{The United States Office of Management and Budget (OMB) classifies counties as metropolitan, micropolitan or neither \cite{hhs}.
A metropolitan statistical area contains a core urban area of at least $50,000$ population.
A metro area represents a functional delineation of an urban area with a network of strong socioeconomic ties, and provision of infrastructure services \cite{stier2022reply, batty2009cities, batty2011cities}.
A micropolitan statistical area contains an urban core of at lest $10,000$ but less than $50,000$ inhabitants. 
}
About $86\%$ of the total U.S. population lives in counties belonging to metro areas representing approximately $28\%$ of the land area of the country. For this reason, our analysis focuses on the growth dynamics of MSA counties.
Supplementary Fig. 2 shows the $3,141$ counties (\textcolor{black}{administrative} subdivisions of the states) in the U. S., comprising about $321$ million inhabitants in the starting year of the \textcolor{black}{ACS 5-Year} survey period ($2015-2019$) of our analysis \cite{ACS}. 


\textcolor{black}{Population growth has two components, namely natural growth and migration.
Natural growth accounts for births minus deaths, and migration comprises of domestic and international migration.
With recent trends showing that births and natural increase have declined in the U.S. and in recent years contribute less to overall city population growth \cite{hamilton2021births, johnson2020births}, migration patterns become more relevant to the study of city population growth.
Because the ACS flow files contain international inflows only, the relative importance of migrations on population growth is here addressed by $x = \rvert$Inflows - Outflows$\rvert / \rvert$Births - Deaths$\rvert$ (Supplementary Figs. 3 and 4), which is the ratio between domestic netflows and natural growth.
}
The statistical distribution of this quantity computed for all U.S. counties is well fitted by a lognormal distribution, and shows that $x \ge 1$ for $76.5\%$ of counties.
For most counties, \textcolor{black}{domestic migration} dominates \textcolor{black}{population growth, and understanding the spatial structure of domestic netflows (and their distribution within a city) is crucial to the comprehension of the mechanisms behind the heterogeneity of city population growth.}

At this spatial granularity, we observe a strong heterogeneity among the U.S. counties (Supplementary Fig. 2) for the period $2015-2019$, along with examples of specific MSAs. 
In particular, \textcolor{black}{the relative dispersion of counties relative growth due to netflows is higher than one for about $85\%$ of the metro areas}, indicating a large heterogeneity within the same city and pointing towards the spatial structure of \textcolor{black}{domestic migration}. The observed difference in the netflows stresses the relevance of our approach: counties belonging to the same city may have specific growth rates \textcolor{black}{due to} population flow patterns, thus indicating preferential flow destinations and pinpointing the direction in which the city has expanded. 

\section{Heterogeneity of \textcolor{black}{inter- and intra-city flows}}\label{sec3}

Inter-\textcolor{black}{city} flows represent the major component of the total flows ($\sim 55\%$), while intra-\textcolor{black}{city} flows represent $\sim 25\%$.
Flows between \textcolor{black}{metro and micro areas, and between metro and non-statistical areas} are the smallest components, with $\sim 13\%$ and $\sim 7\%$, respectively. 
Given that about $80\%$ of the \textcolor{black}{domestic migration} are composed of intra- and inter-\textcolor{black}{city} flows, we will focus \textcolor{black}{our attention on} describing the structure of intra- and inter-\textcolor{black}{city flows}, but in the Supplementary Information we offer a brief analysis of \textcolor{black}{flows between metro and micro areas, and between metro and non-statistical areas}.

Inter-\textcolor{black}{city flows} are not uniform across the U.S. cities. 
The most intense annual netflows ($>2,000$ people per year), \textcolor{black}{accounting for approximately $17\%$ of the entire inter-city U.S. netflows,} are mainly from New York and Chicago to California and Florida (Fig. \ref{map_flows}), and from Los Angeles to neighboring cities. 
Notably, netflows among the Midwestern cities are mostly negative and below the threshold  we set. 
These \textcolor{black}{flows} are mainly responsible for increasing or decreasing the population of a given city. 
\textcolor{black}{Intra-city} flow patterns, illustrated with the $7$ most populous U.S. cities with more than $5$ counties, are also non-uniform. 

Our analysis reveals that city centers (defined as the “core” county with the highest population density) are more likely to have negative netflows, indicating that people are leaving the central \textcolor{black}{regions} of cities.
The arrows in Fig. \ref{map_flows} indicate the direction of the most intense \textcolor{black}{netflows}, \textcolor{black}{supporting} this finding and \textcolor{black}{highlighting} that there is a trend of people \textcolor{black}{moving} from internal to external \textcolor{black}{regions}, contributing to population growth and spatial expansion of U.S. cities. 
\textcolor{black}{
In fact, we found no correlation between relative population growth (viz. population growth by city size) and distance from the core county (Supplementary Fig. 5A) for the $46$ cities with more than $5$ counties, with relative growth about $0.03\pm 0.05$.
On the other hand, we found that relative natural growth (Supplementary Fig. 5B) is negatively correlated with the distance to core county, thus natural growth is less relevant as a component of growth in the outer regions of cities.
Consequently, our results show that not only the contribution of each component of growth changes with distance to core county, but also that the internal redistribution of people is an important mechanism of growth, mainly in the external counties.
}

\begin{figure}[t!]
\centering
\includegraphics[width = 0.6\textwidth]{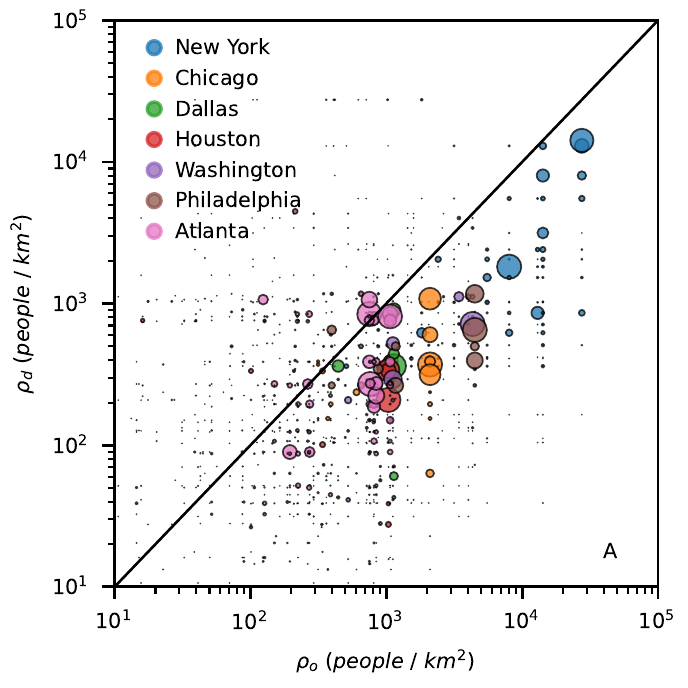}
\includegraphics[width = 0.6\textwidth]{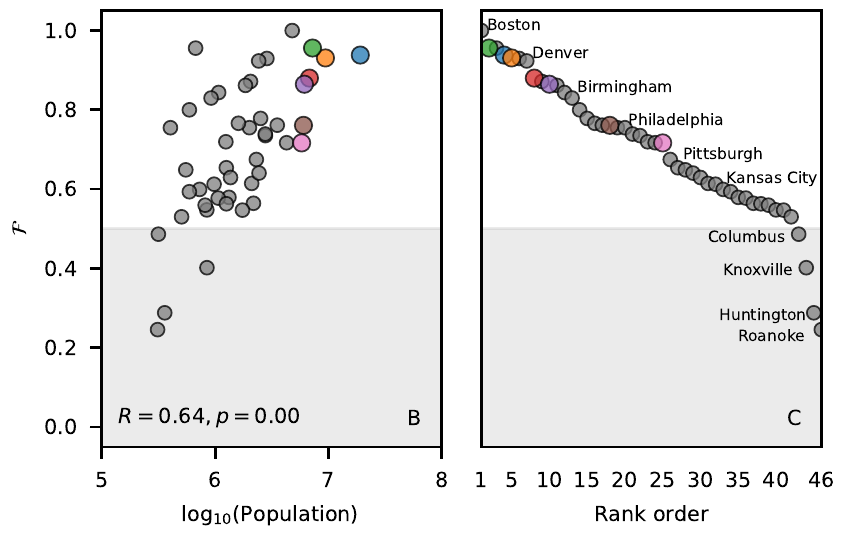}
\caption{People are moving to counties with lower population density. 
(A) The population density of the origin ($\rho_o$) and destination ($\rho_d$) counties of intra-\textcolor{black}{city} netflows for New York, Chicago, Dallas, Houston, Washington D.C., Philadelphia, Atlanta, reveal that the majority of the flows occur from high to low density counties. 
The size of the symbols are proportional to the intensity of the netflow, and the black line corresponds to $y = x$. 
(B) The fraction of netflows to lower density counties $\mathcal{F}$ has a positive correlation with city population when we consider the $46$ MSAs with more than $5$ counties, suggesting that intra-\textcolor{black}{city} netflows to lower density counties are more frequent as the city size increases. 
(C) The ranking of the cities according to $\mathcal{F}$.
}
\label{density_in_out}
\end{figure}

We also examined variability in inter- and intra-\textcolor{black}{city} flows within the $50$ states (Supplementary Fig. 6). 
Total \textcolor{black}{flows} within a state increase, as expected, with the state population. 
Two special cases are, however, of interest: (1) two states (Vermont and Rhode Island) with small populations have only one MSA, in which case within-state inter-\textcolor{black}{city} flows are zero; and (2) nearly $40\%$, or $149$, of MSAs have only one county, in which case intra-\textcolor{black}{city} flows could not be estimated.
For all other cases, we observe on average an equal split between inter- and intra-\textcolor{black}{city} \textcolor{black}{flows}, but with considerable variability among the states, with mean about $0.5$ and standard deviation about $0.2$. 
A generalization of the intra- and inter-\textcolor{black}{city} migratory patterns for all $46$ cities with more than $5$ counties shows that the percentage of migrants from intra- and inter-city flows are of the same order of magnitude (Fig. \ref{flows_with_distance}).

Apart from the core county, \textcolor{black}{flows} from the same city correspond to about $50\%$ of the inflow of people in the counties, presenting a slightly positive correlation with their distance from the city center (Fig. \ref{flows_with_distance}A). 
The low percentage for the \textcolor{black}{core} county indicates that it is not the major destination of flows from the same city. 
The percentage of inflows from other cities is higher in the \textcolor{black}{core} county and decays as we move towards the suburbs (Fig. \ref{flows_with_distance}B). 
The moderate negative correlation of this percentage with the distance reveals that inflows from other cities are more likely to concentrate in the core \textcolor{black}{regions} of a city.

The percentage of outflows directed from the core county to other counties within the same city has a slightly negative correlation with the distance of the origin county to the city center, so it is more likely to find intra-\textcolor{black}{city} flows with outflows from internal \textcolor{black}{regions} (Fig. \ref{flows_with_distance}C). 
The core county is an exception again, suggesting that it is less likely that someone leaving the core
county will move to another county within the same city. 
The slightly negative correlation of the percentage of outflows directed to other cities suggests that there is a trend of people leaving the core
county and the central \textcolor{black}{regions} to move to other cities (Fig. \ref{flows_with_distance}D). 
The high percentage of inflows (Fig. \ref{flows_with_distance}B) and outflows (Fig. \ref{flows_with_distance}D) in the central \textcolor{black}{region} due to inter-\textcolor{black}{city} flows implies that the central \textcolor{black}{regions} of cities are more dynamic and diverse and that people tend to move to counties with similar levels of urbanization. 
The same pattern is observed for flows \textcolor{black}{between metro and micro areas, and for metro and non-statistical areas}, allowing us to conclude that people moving from rural areas are more likely to move to the external \textcolor{black}{regions} of a city (Supplementary Fig. 7).

\begin{figure}[!t] 
\centering
\includegraphics[width = 0.65\textwidth]{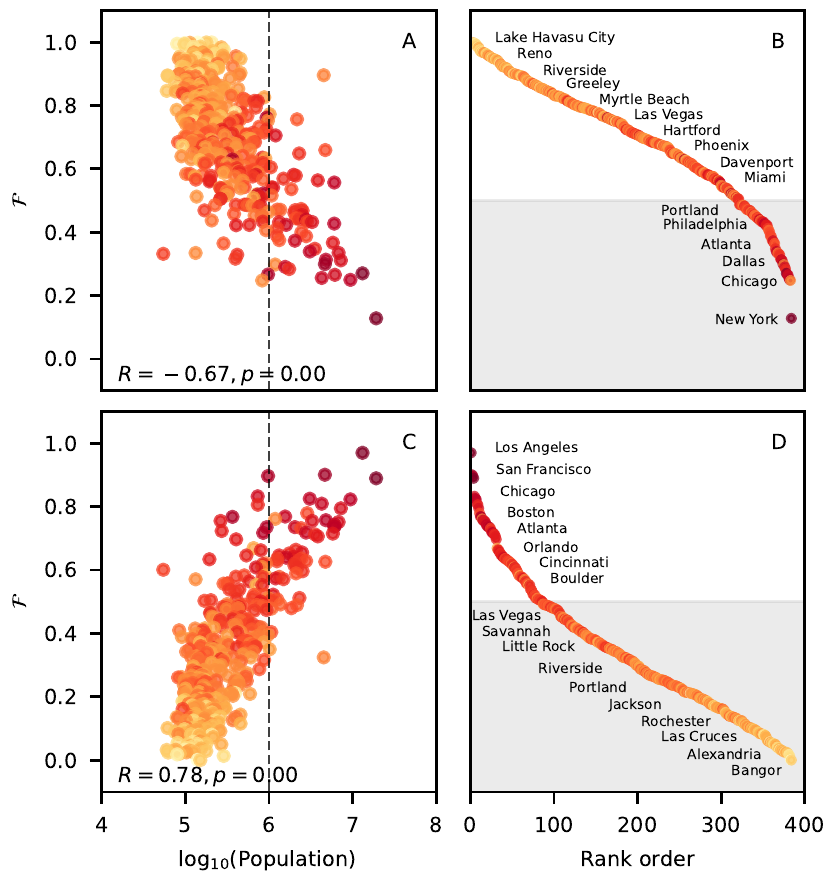}
\caption{Inter-\textcolor{black}{city} flow patterns depend on the population size of the origin and destination cities. 
Each point corresponds to a particular city. 
(A) Fraction $\mathcal{F}$ of netflows going to lower density counties versus the population of the destination city. 
Inflows to counties of \textcolor{black}{large} cities (with population greater than $10^6$, dashed line) usually comes from counties with lower population densities. 
(B) Rank of cities according to the share of inflows from lower density counties. 
(C) Fraction $\mathcal{F}$ versus the population of the origin city. 
Outflows from counties of \textcolor{black}{large} cities usually go to cities with lower density counties. 
(D) The rank of cities according to the share of inter-city netflows to lower density counties is presented.
The dots are colored according to the city population density (darker red means higher density).
}
\label{inter_in_out_flows}
\end{figure}

The positive correlation of the relative growth with the distance due to intra-\textcolor{black}{city} \textcolor{black}{flows} (Fig. \ref{flows_with_distance}E) shows that the resulting \textcolor{black}{intra-city} redistribution of people, given by the difference between inflows and outflows, is such that there is a trend from core county to the external counties (viz. suburbs). 
When compared to the relative growth due to inter-\textcolor{black}{city} flows (Fig. \ref{flows_with_distance}F), which does not show any trend and that have negative values for the most distant counties, it becomes clear that intra-\textcolor{black}{city} \textcolor{black}{flows} play a major role in the \textcolor{black}{population} increase observed in outer \textcolor{black}{regions} of cities. 
\textcolor{black}{Interestingly, large circle and square dots in Figs. \ref{flows_with_distance}E and \ref{flows_with_distance}F suggest that the loss of people due to inter-city netflows is more intense than the gain of people due to intra-city netflows in some external counties of the largest metro areas, thus explaining the population decline in some outer regions of New York and Chicago (as shown in Figs. \ref{map_flows}B and \ref{map_flows}C). 
}

The \textcolor{black}{population growth} due to intra-\textcolor{black}{city} flows is also depicted by Fig. \ref{density_in_out}.
The concentration of flows below the diagonal captures the heterogeneity and the preferential destination of intra-\textcolor{black}{city} \textcolor{black}{netflows}. 
We observe that people are more likely to move to lower population density counties when moving from one place to another within the same city, as exemplified by $7$ cities in panel A. 
Panel B summarizes this analysis for the $46$ cities with more than $5$ counties by showing the fraction $\mathcal{F}$ of intra-\textcolor{black}{city} netflows to lower density counties. 
We note that more than $93\%$ of the cities have $\mathcal{F} > 0.5$ and that there is a positive correlation of $\mathcal{F}$ with the city population, and C shows the rank of cities according to the fraction of intra-\textcolor{black}{city} netflows to lower density counties.

Population density does not seem to play a major role in driving flows between counties of different cities.
The fraction of inter-\textcolor{black}{city} netflows to lower density counties is about $57\%$ when we consider all the $384$ MSAs.
The heterogeneity in the \textcolor{black}{inter-city netflow} pattern can be assessed by analyzing $\mathcal{F}$ versus the population of the destination city (Fig. \ref{inter_in_out_flows}, panels A and B) and $\mathcal{F}$ versus the population of the origin city (Fig. \ref{inter_in_out_flows}, panels C and D).
The negative correlation of $\mathcal{F}$ with the population of the destination city in panel A indicates that inflows are more likely to come from lower density counties as the destination city size increases.
The positive correlation of $\mathcal{F}$ with the population of the origin city in panel C reveals that outflows tend to be directed to lower density counties as the origin city size increases.
The trends observed in panels A and C reveals that inter-\textcolor{black}{city} \textcolor{black}{flows} are more likely between counties with different population densities rather than between counties with similar population densities.
Panels B and D show the rank order of cities according to $\mathcal{F}$ as function of the destination city size and the origin city size, respectively.

\textcolor{black}{We would expect that there might be preferential locations within a given city to which people move due to various factors such as lower costs of housing and employment opportunities.}
\textcolor{black}{
However, it seems that house prices have little to no effect on intra-city netflows (Supplementary Fig. 8).
While the fraction of intra-city netflows to counties with less expensive houses is about $0.8$ for cities like New York, Chicago and Washington, this fraction is about $0.2$ for cities like Dallas, Houston and Philadelphia.
The lack of a clear national pattern highlights the specificity of each city and the heterogeneity of the regional housing market in the U.S. \cite{moench2011hierarchical, aastveit2018asymmetric}. 
On the other hand, the fraction of intra-city netflows to counties with lower unemployment rates is higher than $0.5$ for about $2/3$ of the cities (Supplementary Fig. 9), thus showing that people are more likely to move to counties with lower unemployment rates.
}

\section{Statistical structure of  \textcolor{black}{inter-city flows}}\label{sec4}

Intra-\textcolor{black}{city} flows capture the internal redistribution of population, without altering the total city population. 
In this context, we focus on inter-\textcolor{black}{city} flows to investigate whether or not extreme flows play an important role in shaping the growth of counties as observed at the city level \cite{verbavatz2020growth}.
\textcolor{black}{
For cities, Verbavatz and Barthelemy \cite{verbavatz2020growth} introduce a stochastic equation to describe population growth, composed of two terms.
The first term accounts for out-of-system growth, which includes natural growth and international migration, and the second term accounts for the growth due to domestic netflows.
They find that total netflows adjusted by population size can be well approximated by a Lévy distribution, and this heavy-tailed distribution indicates that  rare and extreme inter-city flows (viz. migratory shocks) dominate city population growth.  
}

\begin{figure}[t!]
\centering
\includegraphics[width = 0.65\textwidth]{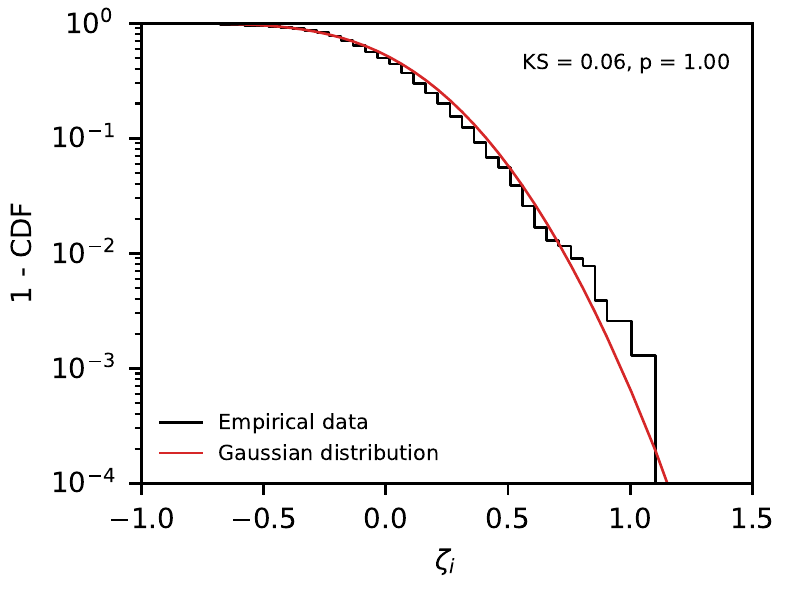}
\caption{Extreme shocks are dissipated at the county level. The distribution of $\zeta_i$, which is the sum \textcolor{black}{of the netflows of a county $i$ adjusted by its population}, suggests that \textcolor{black}{migratory events are exponentially bounded at the county level} since $\zeta_i$ is well described by a Gaussian distribution.
The distribution of $\zeta_i$ is computed here for all the counties with at least $50.000$ inhabitants.
\textcolor{black}{We also show the result of the two-sided $KS$ test under the null hypothesis that $\zeta_i$ follows a Gaussian distribution.} 
}
\label{spilover}
\end{figure}


\textcolor{black}{
Here, we find that, for counties, the distribution of total netflows adjusted by population size, which is represented by $\zeta_i$ and captures the intensity of inter-city migratory flows (see section \ref{methods} for details)}, can be approximated by a Gaussian distribution (Fig. \ref{spilover}).
\textcolor{black}{
The lack of a heavy tail in the empirical distribution of $\zeta_i$ suggests the absence of extreme flows at the county level, thus indicating that the growth of counties can be described by smoother migratory process than cities.
Given that cities do experience migratory shocks \cite{verbavatz2020growth}, our findings indicate that cities redistribute inflows among its different counties, leading to a `spill-over' effect that dampens flow shocks at the county level.
}

\section{\textcolor{black}{Heterogeneity of international inflows}}

\begin{figure}[h!] 
\centering
\includegraphics[width=1\textwidth]{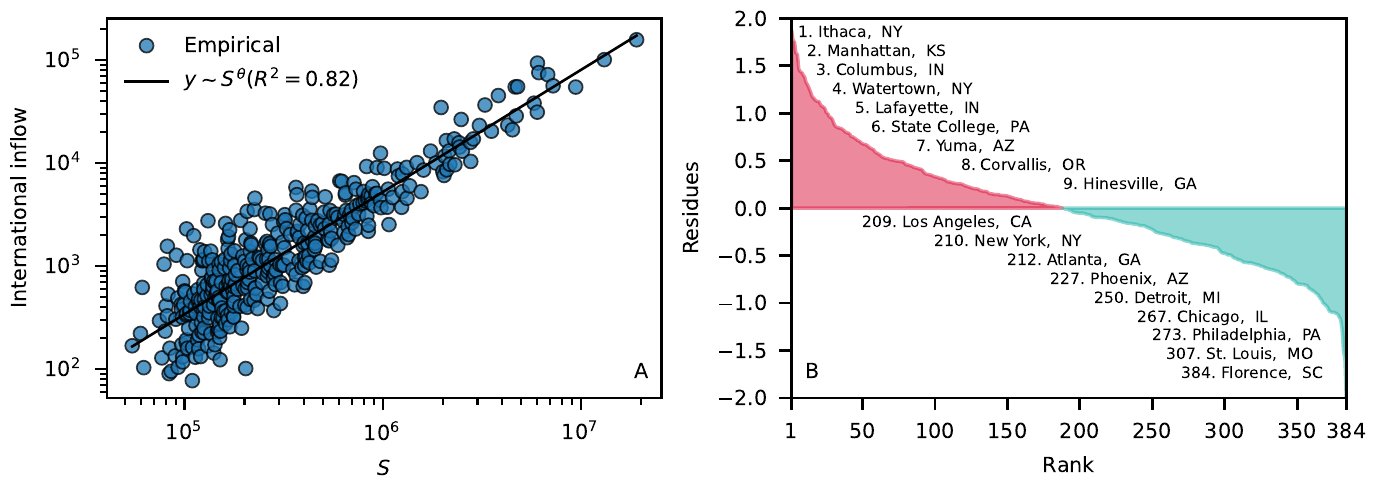}
\caption{
\textcolor{black}{
International inflow scales superlinearly with city size. 
Panel A shows the number of international immigrants as function of the city size $S$ for the $384$ U.S. metro areas. 
Note that the spread of empirical data around the model narrows as the size of the city increases.
Panel B shows the rank of the metro areas and the residues, which capture deviations from the null model thus highligthing cities receiving more/less international inflows than expected.
}}
\label{superlinear_growth}
\end{figure}

\textcolor{black}{
The highest share of international inflows is concentrated in large cities.
About $40\%$ of the international inflows are destined to the top $10$ ($\sim 2.6\%$) largest metro areas of the U.S..
New York is the first with $8.5\%$ of international inflows, followed by Los Angeles and Miami with $5.4\%$ and $5.0\%$, respectively. 
Indeed, international inflows $Y_k$ scale superlinearly with the population $S_k$ of the metro area $k$ (Fig. \ref{superlinear_growth}A), being fitted by the function $Y = Y_0 S^\theta$ in which $\theta = 1.19$ ($95\%$ CI $[1.13, 1.24]$) and $Y_0$ is a normalization constant, thus larger cities have more immigrants per capita than smaller cities.}

\textcolor{black}{
Interestingly, this gain with scale is also observed in socioeconomic city metrics as crime, GDP, innovation and wealth creation due to the manifestation of non-linear agglomeration phenomena \cite{bettencourt2010urban, bettencourt2010unified, bettencourt2013origins}.
Using $Y = Y_0 S^\theta$ as a null model, we can compute deviations from the average behavior by means of residuals given by $\log (Y_k / Y_0 S_k^\theta)$ \cite{bettencourt2010urban}.
The rank of the residues (Fig. \ref{superlinear_growth}B) shows that college towns are among the top metro areas receiving more international inflows than expected, while large cities as Los Angeles, New York, Atlanta, and Chicago are among the metro areas receiving less international inflows than expected.
}

\textcolor{black}{
The spatial distribution of international inflows within cities is shown in Supplementary Fig. 10.
The highest share of inflows is concentrated at core counties, and the percentage of inflows decreases dramatically with the distance from the core county.
This result suggests that inflow of international migrants is an important component of population growth, particularly at the core regions of large cities.
}

\section{\textcolor{black}{Robustness of our findings}}

\textcolor{black}{
Patterns of population redistribution change from time to time in the U.S., and are affected by several factors.
For instance, in the $1960$s non metropolitan counties lost about $3$ million people due to outflows to metropolitan counties, while the reverse trend was observed in the $1970$s when non metropolitan counties experienced net inflows of about $2.6$ million people \cite{brown2013new}.
Wardwell and Brown in \cite{brown2013new} indicate that three factors might be among the main reasons of such change, namely economic decentralization, preference for rural living, and modernization of rural life.
}
\textcolor{black}{
Temporal influence of factors as socioeconomic conditions, transportation infrastructure, natural amenities, and land use and development on population growth in rural and suburban areas is explored in \cite{chi2011population}. 
Changes in rural migration patterns are also studied in \cite{johnson2000continuity}, where age specific rural migration patterns from $1950$ to $1995$ are analyzed. 
In \cite{rayer2001geographic}, the authors explore redistribution trends across U.S. counties from $1980$ to $1995$ split into three five year periods ($1980$-$1985$, $1985$-$1990$, $1990$-$1995$), and \cite{johnson2015migration} analyzes changes in age specific nationwide migration patterns from $1950$ to $2010$.
}

\textcolor{black}{
The spatial structure of migration patterns may indeed change from time to time; our results correspond to the current intra- and inter-city redistribution trends, based on the most recent ACS migration flow files.
We present a thorough empirical and statistical analysis of domestic migration flows among U.S. cities ans counties. Our study also introduces a framework that can be used for analyzing and comparing internal redistribution of people across different time periods. 
Indeed, we extended our analysis for two other time periods, $2005$-$2009$ and $2010$-$2014$.
With respect to the spatial distribution of intra- and inter-city flows, similar trends are observed in both periods (Supplementary Figs. 12 and 13), namely inter-city flows are responsible for the highest share of inflows to core counties, and intra-city flows are responsible for the highest share of inflows to external counties.
We also explored the role of population density in driving netflows between counties within the same metro area in $2005$-$2009$ and $2010$-$2014$.
The results (Supplementary Figs. 14 and 15) indicate that $95.7\%$ of cities were dominated by intra-city moves to lower density counties in $2005$-$2009$, and this percentage dropped to $76.1\%$ in $2010$-$2014$.
Our findings indicate that the trends we report here are taking place since $2005$ but with different intensities.
}

\textcolor{black}{
The robustness of our findings is checked with additional migration data from the Internal Revenue Service (IRS), which reports the year-to-year address changes on individual tax returns filled with the IRS \cite{IRSdata}.
The results obtained with the analysis of IRS datasets from periods $2015-2016$, $2016-2017$, $2017-2018$, $2018-2019$ (Supplementary Figs. 16, 17, 18, 19), reveal similar trends to those we found using ACS data.
Particularly, we observe that, for all periods considered, the correlation between intra-city netflow / $S$ and distance to core county is stronger than we found with ACS data, thus highlighting the role of intra-city flows in driving population to external regions of cities.
The main difference between both datasets is in the percentage of intra- and inter-city inflows and outflows: while ACS data indicates that both flows have approximately the same contribution to the total flows, the IRS data indicates that, besides the core county, intra-city flows are responsible for about $80\%$ of inflows and outflows of metro areas. 
}

\section{Discussion}\label{sec5}


We presented an analysis of the domestic population flows from \textcolor{black}{domestic migration}, disaggregated at the county level, that drive the population growth of U.S. counties and cities. 
We showed that urban population growth is spatially heterogeneous, where intra- and inter-\textcolor{black}{city} flows contribute equally to the population dynamics of cities. Intra-county flows could not be examined in this study (absence of sub-county data), but account for $\sim 60\%$ of all domestic flows in the U.S. (Supplementary Fig. 1). 
Analyzing spatial aspects of these flows is an interesting direction for future studies.

Large polycentric urban agglomerations in the U.S. emerge from the development and merging of several towns and cities with strong socioeconomic ties, even though they retain separate municipal governing structures \cite{louf2013modeling, sarkar2020measuring}. 
Similarly, at the regional-scale, even larger urban agglomerations of multiple cities emerge; \textcolor{black}{ OMB} designated $172$ such agglomerations as Combined Statistical Areas. 
Our analyses of intra-\textcolor{black}{city} flows are based on `core' county, and the spread of data points observed in Fig. \ref{flows_with_distance} reveals the heterogeneity among cities in terms of polycentric organization, but with a primary, central urban county \cite{louf2013modeling, sarkar2020measuring}. 
Recent multi-scale modeling analyses of urban mobility and growth \cite{li2021global, xu2021emergence}; \cite{bettencourt2020urban,barbosa2021uncovering} are noteworthy in combining diverse data sources and theoretical approaches, but there is a need for empirical data analysis at a more disaggregated level \cite{yabe2022toward, yabe2022mobile}.

Counties with the highest population density in MSAs constitute the `core' of cities, characterized by intense inter-\textcolor{black}{city} inflows and outflows. 
Although population growth of cities is shaped by inter-city migratory shocks \cite{verbavatz2020growth}, we showed that counties are not subject to the same population dynamics. 
This result suggests that flow shocks are dispersed among its counties; the `spill-over' effect thus dampens the shocks at the county level. 
The population growth of urban counties through densification is driven by creativity, innovation, and technological advances, but also triggers outflows because of increasing cost of living and decreasing quality of life issues \cite{fan1999vertical, bettencourt2007growth, graham2015luxified, wong2004vertical}. 
Net outflows from the core (most dense) to external counties expand urban sprawl to neighboring counties. 
Inter-\textcolor{black}{city} migrations, and flows from \textcolor{black}{micro and non-statistical areas to metro areas}, are more likely between counties with similar levels of urbanization, showing a preferential flow destination. 

\textcolor{black}{
Not all domestic migration has the same demographic impact. 
For instance, migration of young people might contribute to a larger natural growth.
Migration of elderly people, on the other hand, might have the opposite effect.
Particularly, a good discussion about migration up and down urban hierarchy for different age groups is presented in \cite{plane2005migration}, which helps in understanding the trends and the migration patterns we found.
Plane \textit{et al.} show that the main components of positive population growth in higher density urban settings are international immigration and natural increase since domestic netflows are negative.
Strong outflows towards lower density counties are composed of people in their late $50$s and $60$s preferring less congestion, higher natural amenities, and cheaper housing, thus explaining why natural increase is lower at counties far from the core county (Supplementary Fig. 5B).
Inflows to higher density counties is mainly composed of young, single, and college-educated adults in the $25$-$29$ year age group.
Interestingly, once they reach their mid-career stage and start their family, the migration trends are reversed: we observe trends towards lower density counties for $30$-$34$ and $35$-$39$ year age groups, mainly prompted by housing costs, school quality and suburban road congestion \cite{plane2005migration}.
}

\textcolor{black}{
In this context, we propose a framework for studying the spatial distribution of migratory flows.}
We have focused here on the U.S. because of the public availability of robust data sets at the county level, but our conclusions could be extended to other developed and highly urbanized countries where
\textcolor{black}{the highest share of domestic migration is composed of intra- and inter-city flows.}
\textcolor{black}{
However, the general intra- and inter-city flow patterns we reported here might not hold for low and middle income countries presenting lower urbanization levels.
For instance, city population growth in countries experiencing rapid urbanization in Asia and Sub-Saharan Africa is mainly composed of rural-to-city flows, in which international migration of refugees and the emergence of large informal settlements are also important components of urban growth \cite{potts2013rural, cockx2017corn, saghir2018urbanization, ostby2016rural, lohnert2017migration, zhang2003rural, malik2017major, soman2020worldwide}.
}

\textcolor{black}{
Metropolitan statistical areas are surrounded by rural counties.
As the population of the adjacent regions increase and cities expand, rural counties are reclassified as metro counties.}
\textcolor{black}{
The reclassification of the external counties as urban places in the U.S. fails to recognize that most of the population might in fact be rural \cite{lichter2021rural}. A recent global-scale analysis \cite{cattaneo2021global} suggested that about a quarter of the global population lives in periurban areas of intermediate and small cities.  
Interestingly, they find that the highest share of population is found at large cities and proximate areas for high income countries, whereas low income countries have the highest share of population concentrated in small cities and proximate areas.
Migration data are needed in order to construct a typology of flows in different parts of the world.
}

\textcolor{black}{
In this paper, we have studied the mechanisms behind the heterogeneous growth of cities.
Growth of cities leads to many benefits, which serve to increase the attractiveness of the city.
As cities grow, wealth and innovation per capita increases since these quantities scale superlinearly with city size as a result of agglomeration effects \cite{bettencourt2007growth}. 
Simultaneously, the volume occupied by infrastructure scales sublinearly with city size, and this economy of scale means that large cities need less infrastructure per capita than small cities \cite{bettencourt2013origins}.}
\textcolor{black}{
Conversely, land-use changes from expansion of metro areas also has costs at local, regional and national scales.
Such costs include, among others, loss of agricultural areas (food security) \cite{bren2017future}, fragmentation of natural areas (loss of ecosystem services) \cite{li2022global, liu2016relationship}, and growing resource demands extracted from increasingly remote locations (impacts on ecosystem impacts at larger scales) \cite{yeh2012global}. 
Heterogeneity of cities both manifests and amplifies socioeconomic inequality, thus contributing diminished urban community resilience.
}

Understanding the spatial organization of the migratory flows is a step towards understanding the drivers of urban growth and heterogeneities in cities. We found that it is necessary to distinguish these flows into different components according to their destination (central vs. external county) and these flows are probably governed by different underlying reasons and household cost-benefit analyses. The possibility of constructing a typology of flows allows then to test the influence of various parameters and models, to analyze the dynamics of inequalities within cities, and eventually to help urban planners to forecast urban expansion and densification.

\section{Methods}\label{sec6}

\subsection{\textcolor{black}{Data collection}}

\textcolor{black}{
Our analysis is focused on a five year period, from $2015$ to $2019$. 
We used two main sources of data sets, both from the U.S. Census. 
The first main source is the ACS County-to-County Migration Files \cite{ACS}.
Annually, approximately $3.54$ million independent housing units addresses were selected among all the U.S. counties.
There were four modes of data collection: internet, mail, telephone, and personal visit, in which respondents are asked whether they lived in the same residence one year ago.
}
\textcolor{black}{
The results, reported over $5$-year periods for robust flow estimates of less populated counties, are made available cleaned and pre-processed in spreadsheet files \cite{ACS}.
From this file, we obtain estimates of inflows and outflows between pairs of counties, in which ``flow estimates resemble the annual number of movers between counties for the $5$-year period data was collected'' \cite{ACS2}.
}

\textcolor{black}{The second main source is the County Population Totals: 2010-2019 dataset \cite{CPT}, which offers ``population, population change, and estimated components of population change'' from April 1, 2010 to July 1, 2019.
Using the resident population from the $2010$ Census as a starting point (population base), county population estimates are derived from the following demographic balancing equation: population estimate = population base + births - deaths + migration \cite{CPT} (see \cite{CPT} for detailed explanations of how births, deaths and migration are estimated).
From this data set, which is made available cleaned and pre-processed in a spreadsheet table, we obtain the domestic netflows, births and deaths for each county from July 1, 2015 to June 30, 2019 used to compute the quantity $x$.
}
We focus our study on $3,141$ counties and $384$ U.S. metro areas.

\textcolor{black}{
To analyze the housing prices of origin and destination counties, we used the housing data from Zillow Research \cite{Zillow}.
Zillow publishes the Zillow Home Value Index, which reflects the typical values of homes across the U.S..
The data is also made available at the county level, cleaned and pre-processed in a spreadsheet table.
}

\textcolor{black}{
The IRS data we used to check the robustness of our findings were collected from the IRS - SOI Tax Stats - Migration Data website \cite{IRSdata}.
From \cite{IRSdata}: ``Migration data for the United States are based on year-to-year address changes reported on individual income tax returns filed with the IRS."
Data files are made available for download cleaned and preprocessed, in comma separated values files (.csv file extension).
}

\subsection{\textcolor{black}{Analyzing inter-city flows}\label{methods}}

\textcolor{black}{
Here, we show that the distribution of normalized netflows at the county level is exponentially tempered, suggesting that counties do not experience extreme migratory events as do cities. 
}
At the county level,
\textcolor{black}{
we define $J_{i, k}$ as the aggregate flow from county $i$ to all counties of MSA $k$, and $J_{k, i}$ as the aggregate flow from all the counties of MSA $k$ to county $i$.
Following \cite{verbavatz2020growth}, we assume that $J_{i,k} = I_0 S_i^\mu S_k^\nu x_{i, k}$, in which $I_0$ is a constant, $S_i$ is the population of county $i$, $S_k$ is the total population of MSA $k$, $\mu$ and $\nu$ are exponents of $S_i$ and $S_k$, respectively, and $x_{i, k}$ accounts for random noises and higher order effects.
In this notation, the total netflow of county $i$ is given by $\mathcal{J}_i = \sum_{k \in N_i}{(J_{i, k} - J_{k, i}})$, where $N_i$ is the set of MSAs exchanging people with county $i$.
}

\textcolor{black}{
In order to reduce the number of free parameters in the expression for $J_{i, k}$, we define the flow per capita $I_{i, k} = J_{i, k} / S_i$.
Given that the ratio $I_{k, i} / I_{i, k} = (S_i / S_{k})^{\nu - \mu + 1}$ can be written as a linear function of $S_i / S_{k}$ (see Supplementary Fig. 11A), we obtain $\nu = \mu$. 
Fitting the flow per capita $I_{i,k}$ versus ${S_i^\nu}{S_{k}^{\nu - 1}}$ gives us $\nu = 0.34$ ($95\%$ CI $[0.33, 0.35]$, Supplementary Fig. 11B). 
As seen in \cite{verbavatz2020growth}, migratory shocks can be captured by the quantity $X_{i, k} = (J_{i, k}-J_{k, i}) / {I_0}{S_i^\nu}$, which measures the relative magnitude netflows with respect to the county population.
Interestingly, the variable $\zeta_i = (1/N_i) \sum_{k \in N_i}{X_{i, k}} = \mathcal{J}_i / I_0 N_i S_i^\nu$, which is the relative impact of the sum of all netflows in county $i$ and captures the intensity of migratory shocks}, can be approximated by a Gaussian distribution (Fig. \ref{spilover}).

\backmatter

\bmhead{Supplementary information}

Supplementary Information is available for this paper.


\bmhead{Acknowledgments}

PSCR was supported, in part, by the Lee A. Reith Endowment in the Lyles School Civil Engineering at Purdue University.
\textcolor{black}{
We would like to thank the three anonymous reviewers for their useful comments, which allowed us to improve the quality of the manuscript.
}




\bibliography{0_REFERENCES}


\end{document}